\documentclass[twoside]{report}
\usepackage{iwsm}
\usepackage{graphicx}

\usepackage{latexsym}
\usepackage{amsfonts}  %to define mathematical symbols, e.g. \Box
\usepackage{amsmath}

\begin{document}

\title{A statistical model for the relation between exoplanets and their host stars}

%({\small \tt Submitted for poster presentation}) %

\titlerunning{A statistical model for exoplanets}

% Authors and running list of authors to be used as right header:
\author{ E. Mart\'inez-G\'omez\inst{1}\inst{,}\inst{2}
and G. J. Babu\inst{2}  }
\authorrunning{Mart\'inez-G\'omez and Babu}
% Full address for correspondence of all authors:
\institute{Instituto de Astronom\'ia, Universidad Nacional Aut\'onoma de M\'exico, Apdo.
Postal 70-264, Ciudad Universitaria, 04510, M\'exico D. F., M\'exico
\and Center for Astrostatistics, 326 Thomas Building,
The Pennsylvania State University, University Park, PA, 16802-2111, USA

{\tt Contact and presenting author:affabeca@gmail.com}
 }

\abstract{A general model is proposed to explain the relation
between the extrasolar planets (or exoplanets) detected until June
2008 and the main characteristics of their host stars through
statistical techniques. The main goal is to establish a mathematical
relation among the set of variables which better describe the
physical characteristics of the host star and the planet itself. The
host star is characterized by its distance, age, effective
temperature, mass, metallicity, radius and magnitude. The exoplanet
is described through its physical parameters (radius and mass) and
its orbital parameters (distance, period, eccentricity, inclination
and major semiaxis). As a first approach we consider that only the
mass of the exoplanet is being determined by the physical properties
of its host star. The proposed model is then validated through statistical analysis.}


\maketitle

\section{Introduction}
An extrasolar planet (or exoplanet) is a planet which orbits a star
other than the Sun, and therefore belongs to a planetary system
other than our Solar System. The first extrasolar planet around a
main sequence star was discovered in 1995 (Mayor and Queloz, 1995).
Actually more than 300 exoplanets have been documented and most of
them with masses greater than Jupiter's mass (Schneider, 2009).
Detecting an exoplanet is a very difficult task because they do not
emit any electromagnetic radiation of their own and are completely
obscured by their extremely bright host stars, that is, normal
telescope observation techniques cannot be used. Thus, in order to
find exoplanets, a variety of techniques like the radial velocity,
pulsar timing, astrometry, gravitational lensing, spectrometry and
photometry (De Pater and Lissauer, 2001) are used. The main purpose
of any method is to detect the effect produced by the exoplanet on
its stellar system. Besides the discoveries it is important to
search for models that can explain the origin, formation and
possible migration of these bodies. For example, Rice and Armitage
(2005) have investigated how the statistical distribution of
extrasolar planets may be combined with knowledge of the host stars'
metallicity to yield constraints on the migration histories of gas
giant planets. Moreover in a series of papers (Udry et al., 2003;
Santos et al., 2003; Eggenberger et al., 2004; Halbwachs et al.,
2005) the emerging properties of planet-host stars and
characteristics of the different orbital-element distributions of
exoplanetary systems have been studied. In this work we analyze the
cross-sectional data for the exoplanets detected until June 2008
through linear regression techniques. The purpose of this kind of
analysis is to verify the relation between the host star and its
orbiting planet. For example, if the planet's mass is strongly
determined by the type of star and hence affects the planetary
formation stage.

\subsection{Characteristics of the data catalog: Stars and Planets}
The catalog was created in February 1995 to facilitate the progress
of the new field named Exoplanetology through the publication of
recent detections and their associated data. The catalog is
interactive and it is available in the webpage: {\sl{http://exoplanet.eu}}.

Until June 2008 the catalog contains: 303 exoplanets and 259
planetary systems (31 multiple systems). Two important
considerations are: 1) the mass of the exoplanet is -at least- 13
$M_J$ (Jupiter's mass) and 2) the data source must be reliable, that
is, previously published in referred journals, presented in
conferences, among others.
 \begin{itemize}
  \item {\bf Stars:} The stellar data are taken from well-known
  databases like Simbad or directly from published papers. The basic
  physical characteristics of a star are: radial velocity, mass,
  metallicity, age and distance.

  \item {\bf Planets:} These data are taken from published papers
  and from the sites: Anglo-Australian Planet Search; California and
  Carnegie Planet Search; Geneva Extrasolar Planet Search Programmes;
  Transatlantic Exoplanet Survey and the
  Department of Astronomy at University of Texas.

\end{itemize}

\section{The General Model: Multiple Regression Analysis}
We start with the following model (Model A) described by the equation:
\begin{equation}
M_{P}=\alpha_{1}+\alpha_{2}DS+\alpha_{3}AS+\alpha_{4}TS+\alpha_{5}MS+\alpha_{6}METAL+\alpha_{7}MAG+\alpha_{8}RS+u_{i}
\end{equation}

\vspace{5mm}

where $M_P$ is the exoplanet's mass and $\alpha_{i}$ are the
coefficients for each term. Eq. (1) expresses the exoplanet's mass $M_{P}$
in terms of the values of the variables representing the features of the host star.
This set of variables contains: the distance, $DS$; the
age, $AS$; the temperature, $TS$; the mass, $MS$; the metallicity,
$METAL$; the magnitude, $MAG$ and the radius, $RS$. Finally $u_{i}$
are the random errors.

We estimate the unknown parameters in Eq. (1) by Ordinary Least
Squares (OLS). The results are shown in Table 1 where we also
include the values of the t-statistics and their associated
probabilities for the coefficient significance tests. From the
estimated values we conclude that the only significant variable for
the Model A is $RS$.

\begin{table}
\centering \caption{\label{tab:1} Estimated values for the
parameters.}
\vspace{3mm}

\begin{tabular}{c c c c c}
\hline\hline
\multicolumn{5}{c}{\bf{Model A}}\\
Variable&$\alpha_{i}$&Standard
Error&t-statistic&Probability\\[0.5ex]
\hline

\sl{C}     & -9.1773  & 5.0051   & -1.8336   & 0.0685 \\
\sl{DS}    & -0.0113  & 0.0074   & -1.5215   & 0.1301 \\
\sl{ES}    & 0.0288   & 0.0872   & 0.3299    & 0.7419 \\
\sl{TS}    & 0.0013   & 0.0007   & 1.9169    & 0.0570 \\
\sl{MS}    & 1.9385   & 1.3721   & 1.4128    & 0.1596 \\
\sl{METAL} & -1.7493  & 1.2586   & -1.3899   & 0.1664 \\
\sl{MAG}   & 0.3689   & 0.2850   & 1.2944    & 0.1973 \\
\sl{RS}    & 0.2335   & 0.1183   & 1.9747    & 0.0499 \\
\hline\hline

\multicolumn{5}{c}{\bf{Model B}}\\
Variable&$\alpha_{i}$&Standard
Error&t-statistic&Probability\\[0.5ex]
\hline

\sl{C}           & -2.5169  & 1.0840   & -2.3218  & 0.0213 \\
\sl{ES}          & -0.0493  & 0.0321   & -1.5345  & 0.1265 \\
\sl{TS}          & 0.0003   & 0.0002   & 1.7417   & 0.0831 \\
\sl{MS}          & 1.1772   & 0.3964   & 2.9698   & 0.0033 \\
\sl{METAL}       & -1.1370  & 0.5001   & -2.2738  & 0.0241 \\
\sl{SIST*MS}     & -0.1809  & 0.2188   & -0.8269  & 0.4093 \\
\sl{SIST*METAL}  & 0.5629   & 0.8338   & 0.6752   & 0.5003 \\
\hline
\end{tabular}

\end{table}

\subsection{Verification of the linear regression assumptions (Model A)}
\begin{enumerate}
\item {Linearity:} The model passed all the Ramsey tests for linearity. We conclude
that the proposed functional form is adequate.
\item {Omitted Variables:} According to the star formation theory, the variables $MS$ and $METAL$ must be included to explain
the relation between the mass of the exoplanet and its host star.
\item {Multicollineality:} There is a possible weak correlation between $MAG$ and $DS$.
\item {Heteroskedasticity:} From the White test on the residuals we conclude that they are not
heteroskedastic, that means the residuals are homoskedastic.
\item {Normality:} From the value of the Jarque-Bera statistic we conclude that the residuals are not
normally distributed.
\item {Homogeneity:} Defining the "dummy" variable as $SIST$ (0 means that the exoplanet belongs to a
single planetary system and 1 refers to a multiple planetary system)
we conclude that the model is homogeneous.
\end{enumerate}

Statistical model must satisfy all the assumptions mentioned
above to be correctly specified. In our case, the
Model A needs some modifications, for example, another functional
form and/or the consideration of an adequate "dummy" variable. In
such a case we derive the Model B:

\begin{equation}
log(M_{P})=\alpha_{1}+\alpha_{2}ES+\alpha_{3}TS+\alpha_{4}MS+\alpha_{5}METAL+\gamma_{1}SMS+\gamma_{2}SMET+u_{i}
\end{equation}

\vspace{5 mm} where $SMS=SIST*MS$ and $SMET=SIST*METAL$ are two new
variables that take into account the fact that the exoplanet can
belong to a single or a multiple planetary system. The parameters are
estimated through OLS and the results are summarized in
Table 1.

\subsection{Verification of the linear regression assumptions (Model B)}
\begin{enumerate}
\item {Linearity:} The model passed all the Ramsey tests for linearity. Moreover we conclude
that the new functional form is more adequate than the presented in
Model A.
\item {Omitted Variables:} The tests indicate that the variables $ES$ and $TS$ must be excluded. However, under this situation
the linearity is not preserved and we loose important physical
information about the host star.
\item {Multicollineality:} There is no correlation among the independent variables.
\item {Heteroskedasticity:} From the White test on the residuals we conclude that they are heteroskedastic.
\item {Normality:} From the value of the Jarque-Bera statistic we conclude that the residuals are
normally distributed.
\item {Homogeneity:} The model has already included the effect of a dummy variable.
\end{enumerate}

Model B ({\sl{log-linear}}) is slightly better than Model A in the
sense that we have improved some of the discrepancies previously
detected in the basic assumptions. However, this latter model cannot
be considered yet to explain the relation between an exoplanet and
its host star. Including the effect of a "dummy" variable seems to
be a clue for another type of model. This binary behavior is
discussed in the next section.

\section{Multiple Regression Analysis with Binary Dependent Variables: a different approach}

Based on the data, the dependent variable (exoplanet) is
simultaneously determined by several parameters, qualitative and
quantitative. In this work we have just assumed that the mass,
$M_P$, is the quantitative variable that represents the whole
physical/orbital characteristics of the planet. However, this fact
is not completely true and more qualitative information must be
taken into account for the model.

In the context of the variable {\sl{exoplanet}}, the relevant
information can be captured by defining a binary variable or a
zero-one variable. An example of such a variable was introduced in Section 2 as
$SIST$ and it is related to the fact that the exoplanet can belong
to a single or a multiple planetary system, in other words, $SIST=0$
if the exoplanet belongs to a single planetary system and $SIST=1$
in other case.

Under this new approach some binary models can be employed and their
choice depends on the data distribution. For example, for a normal
distribution we apply the {\sl{probit}} model, for a logistic
distribution we apply the {\sl{logit}} model and when the data are
truncated or censored we apply the {\sl{tobit}} model.

Once the model is selected, its parameters can be estimated through the
traditional methods like the Maximum Likelihood (ML) and Ordinary
Least Squares (OLS).

A general binary model (Model C) for this case can take the form:

\begin{equation}
M_{P}=\alpha_{1}+\alpha_{2}ES+\alpha_{3}TS+\alpha_{4}MS+\alpha_{5}METAL+u_{i}
\end{equation}

The special case of Model C under the binary context will be discussed elsewhere.
Recently a {\sl{logit}} model was developed and validated by Fressin et al. (2009).
In that work the authors performed a logistic regression to model the probability
that a given planet is "real" (that means, observed or detected) or
just simulated.

\section{Summary and Conclusions}
From our extensive statistical analysis we conclude that Model B is
better than Model A. We have improved its specifications through the
deletion of variables like $MAG$, $DS$ and $RS$ and the addition of
new ones that consider the possibility of finding exoplanets in
single or multiple planetary systems. At the moment this is our best
representation of the relation between the exoplanet and its host
star and in a future work we will consider the problem by including
binary variables.

 \vspace{5mm}

{{\small {\sl{Acknowledgements}}

E. Mart\'inez-G\'omez thanks to DGAPA-UNAM postdoctoral fellowship
and to the Faculty For The Future Program for the financial support provided for this work.
G. Jogesh Babu is supported in part by a National Science Foundation grant
AST-0707833.}

{\small
\section{References}
\begin{description}
\item[De Pater, I. and Lissauer, J. J.] (2001).
{\sl Planetary Sciences}, Cambridge University Press, Chapter 13,
576 pp.

\item[Eggenberger,A., Udry, S. and Mayor, M.] (2004). Statistical properties
of exoplanets. III.
Planet properties and stellar multiplicity. {\sl Astronomy and
Astrophysics}, {\bf 417}, 353-360.

\item[Fressin, F., Guillot, T. and Nesta, L.] (2009). Interpreting
the yield of transit surveys: Are there groups in the known
transiting planets population?. {\sl Astronomy and Astrophysics}, in
press.

\item[Halbwachs,J. L., Mayor, M. and Udry, S.] (2005). Statistical properties
of exoplanets. IV.
The period-eccentricity relations of exoplanets and of binary stars.
{\sl Astronomy and Astrophysics}, {\bf 431}(3), 1129-1137.

\item[Mayor, M. and Queloz, D.] (1995). A Jupiter-mass companion to a
solar-type star. {\sl Nature}, {\bf 378}, 355-359.

\item[Rice, W. K. M. and Armitage, P.J.] (2005).
Quantifying Orbital Migration from Exoplanet Statistics and Host
Metallicities. {\sl The Astrophysical Journal}, {\bf 630}(2),
1107-1113.

\item[Santos, N. C., Israelian, G., Mayor, M., Rebolo, R. and Udry, S.] (2003).
 Statistical properties of exoplanets. II.
 Metallicity, orbital parameters, and space velocities.
{\sl Astronomy and Astrophysics}, {\bf 398}, 363-376.

\item[Schneider, J.] (2009).
Interactive Extra-Solar Planets Catalog. ({\sl
http://exoplanet.eu}).

\item[Udry, S., Mayor, M. and Santos, N. C.] (2003). Statistical properties of
exoplanets. I.
The period distribution: Constraints for the migration scenario.
{\sl Astronomy and Astrophysics}, {\bf 407}, 369-376.

\item[Wooldridge, J. M.] (2008).
{\sl Introductory Econometrics: A Modern Approach}, South-Western
College Pub., 4th ed., 865 pp.

\end{description}
}
\end{document}